\documentclass[pra,showpacs,groupedaddress,amssymb,twocolumn,notitlepage,nofootinbib,longbibliography,floatfix,superscriptaddress]{revtex4-1}
\usepackage{graphicx,amsmath,dsfont}
\usepackage[dvipsnames]{xcolor}
\usepackage{subcaption}
\usepackage{comment}
\usepackage{enumerate}
\usepackage{float}
\usepackage{hyperref}
\hypersetup{
    colorlinks=true,
    linkcolor=red,
    filecolor=magenta,      
    urlcolor=blue,
    pdftitle={Overleaf Example},
    pdfpagemode=FullScreen,
    }
\usepackage{color}
\usepackage{amsthm}
\usepackage{amsmath,amssymb}

\usepackage{xcolor}
\pagecolor{white}
\usepackage[english]{babel}
\usepackage[normalem]{ulem}

\usepackage{calrsfs}

\usepackage{animate}

\newcommand{\lettersection}[1]{\emph{#1.---}}

\newcommand{\eea}{\end{eqnarray}}
\newcommand{\bea}{\begin{eqnarray}}
\newcommand{\ee}{\end{equation}}
\newcommand{\be}{\begin{equation}}
\newcommand{\beq}{\begin{equation}}
\newcommand{\eeq}{\end{equation}}
\newcommand{\beqnn}{\begin{equation*}}
\newcommand{\eeqnn}{\end{equation*}}
\newcommand{\bes} {\begin{subequations}}
\newcommand{\ees} {\end{subequations}}

\AtBeginDocument{%
    \newwrite\bibnotes
    \def\bibnotesext{Notes.bib}
    \immediate\openout\bibnotes=\jobname\bibnotesext
    \immediate\write\bibnotes{@CONTROL{REVTEX41Control}}
    \immediate\write\bibnotes{@CONTROL{%
    apsrev41Control,author="08",editor="1",pages="1",title="0",year="1"}}
     \if@filesw
     \immediate\write\@auxout{\string\citation{apsrev41Control}}%
    \fi
}%

\begin{document}
\title{Meeting the Needs of the Global Quantum Science Community: A Call to Action}

\author{Tzula B. Propp}\altaffiliation{These authors contributed equally and are corresponding authors: propp@physics.leidenuniv.nl; btodd@spicescience.org}\affiliation{Diversity in Quantum (DiviQ), DiviQ.org}
\affiliation{Quantum Flagship Equity, Diversity, and Inclusion Working Group}
\affiliation{Leiden Institute of Physics\\ Leiden University, 2333 CA Leiden, The Netherlands}
\affiliation{SPICE Science, spicescience.org\\  1371 East 13th Avenue, Eugene, OR, 97403, United States}
\affiliation{We in Quantum Development (WIQD), University of Amsterdam, \\
Sciencepark 904, 1098 XH Amsterdam, The Netherlands}
\author{Brandy Todd}\altaffiliation{These authors contributed equally and are corresponding authors: propp@physics.leidenuniv.nl; btodd@spicescience.org}
\affiliation{SPICE Science, spicescience.org\\  1371 East 13th Avenue, Eugene, OR, 97403, United States}
\author{Sara A. Metwalli}\altaffiliation{smetwall@ed.ac.uk}\affiliation{Diversity in Quantum (DiviQ), DiviQ.org}
\affiliation{Quantum Software Lab, School of Informatics\\
University of Edinburgh
10 Crichton Street, Edinburgh, EH8 9AB, UK}
\author{Alina Helena S. Gallardo}\altaffiliation{alina.sanchez@oulu.fi}\affiliation{Diversity in Quantum (DiviQ), DiviQ.org}
\author{Michael Dascal}\affiliation{Diversity in Quantum (DiviQ), DiviQ.org}
\author{Denise Ruffner}\affiliation{Diversity in Quantum (DiviQ), DiviQ.org}
\author{Klaus D. Jöns}
\affiliation{Quantum Flagship Equity, Diversity, and Inclusion Working Group}
\affiliation{ Institute for Photonic Quantum Systems (PhoQS), Center for Optoelectronics and Photonics Paderborn (CeOPP) and Department of Physics, Paderborn University, Paderborn, Germany.}
\author{Shaeema Zaman}
\affiliation{Quantum Flagship Equity, Diversity, and Inclusion Working Group}
\affiliation{Science Melting Pot, Universitetsbyen 14, 8000 Aarhus, Denmark}
\author{Judith Kreukels}
\affiliation{We in Quantum Development (WIQD), University of Amsterdam, \\
Sciencepark 904, 1098 XH Amsterdam, The Netherlands}
\author{Marilù Chiofalo}
\affiliation{Physics Department,
University of Pisa \\
Largo Bruno Pontecorvo, 3
56126 Pisa, Italy}
\author{Lydia Sanmartí-Vila}
\affiliation{Quantum Flagship Equity, Diversity, and Inclusion Working Group}
\affiliation{ICFO-Institut de Ciències Fotòniques, The Barcelona Institute of Science and Technology, 08860, Castelldefels, Barcelona, Spain}

\begin{abstract}

\end{abstract}

\maketitle

\section{Introduction}

$2025$ marks one hundred years since the discovery of quantum mechanics. In the century since then, quantum science has blossomed into a global community composed of academics, engineers, developers, and entrepreneurs. The world is currently in the middle of the so-called second quantum revolution \cite{dowling}, with increased public awareness of quantum science and technology \cite{QuantumFlagship2025}, and growing investment \cite{McKinseyQuantumInvestment} in both quantum hardware and software applications. However, representation remains low among historically marginalized groups: women, LGBTQ+, BIPOC, and people from the global south make up disproportionately few physicists (see e.g. Ref. \cite{Hennessey2024}). There are numerous efforts to improve diversity within quantum science \cite{Heidt2024}, including through workforce development \cite{Ivory2023}. But many of the changes enacted at the highest levels have failed to result in real change, as highlighted and discussed in the recent Women For Quantum Manifesto of Values \cite{Beige2025}. Here, we at Diversity in Quantum and the European Union Quantum Flagship Equity, Diversity, and Inclusion (EDI) Working Group seek to echo and amplify the need for real change in the quantum ecosystem, emphasizing intersectionality and a feminist approach that centers the most vulnerable members of the quantum community: young students and researchers, especially those communities historically marginalized from quantum science.

This report is our attempt to help quantum communities meet this need; we have conducted a survey of quantum scientists all over the world, and here we include both a preliminary report of our findings and  policy suggestions we have built to address them. The primary results of our survey are that, 1) marginalized quantum scientists are experiencing hardships and challenges more than their more privileged peers across all metrics, 2) that this fact is hurting retention of diverse, talented quantum scientists in our field, and 3) quantum EDI is an investment in talent retention and resilience building, which are essential for a thriving, globally competitive quantum ecosystem.

\begin{figure}[t]
    \centering
    \includegraphics[width=\linewidth]{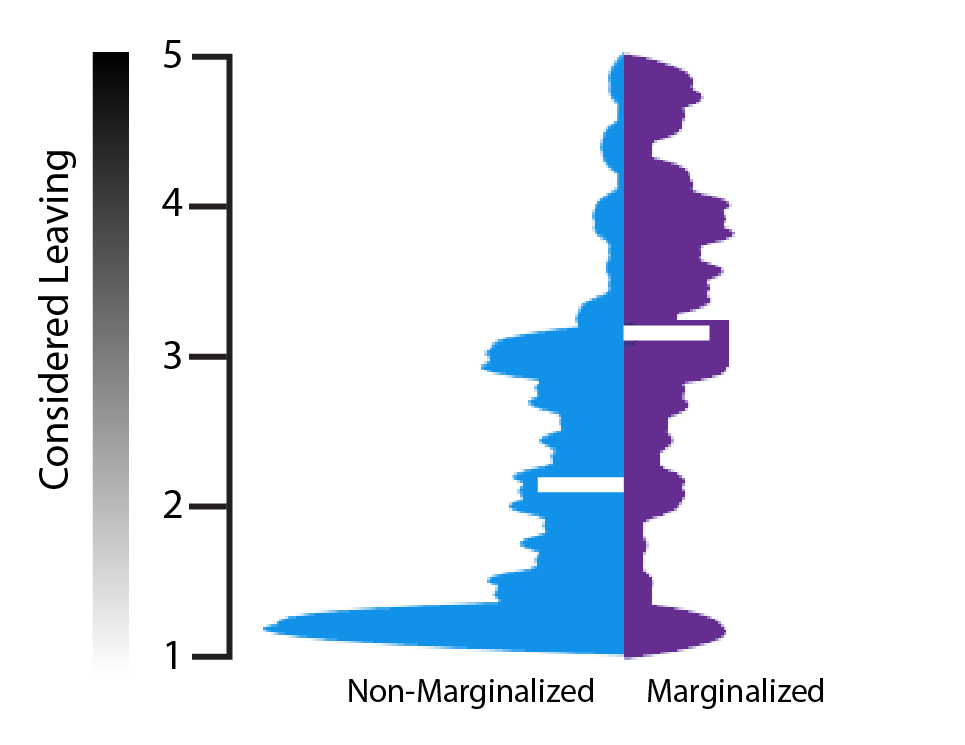}
    \caption{ Mean differences between those who considered leaving quantum science non-marginalized ($\textit{M}=2.05,\, \textit{SD}=1.01$), and marginalized respondents on a scale of 1-5 with 5 being the respondent strongly considered leaving quantum science ($\textit{M}=3.12, \,\textit{SD}=1.19, \,\textit{t}[347]=-9.06,\,\textit{p}<.001$). Notably, very few non-marginalized quantum scientists considered leaving the field compared with over half of marginalized quantum scientists surveyed.}
    \label{fig:leaving}
\end{figure}

\section{Who We Are}

Diversity in Quantum (DiviQ) is an international consortium of researchers, educators, and industry professionals committed to broadening participation, improving workforce climate, and strengthening equitable pathways in quantum science and technology. In partnership with the Quantum Flagship's Working Group for Equity, Diversity, and Inclusion, DiviQ works to translate research on diversity and inclusion into evidence-driven action for the quantum ecosystem. The breadth of our membership gives us a unique vantage point: we see how quantum careers begin, how they stall, and what supports are missing. We also see enormous enthusiasm from early-career scientists who want a quantum field that is not only world-leading, but world-welcoming.

The Quantum Flagship Equity, Diversity, and Inclusion Working Group (EDI-WG) is a collection of individuals from the European quantum community in academia, industry, and governance. Together, we are working to promote a better quantum ecosystem in Europe and beyond with equitable representation and diverse participation through advocacy, surveying, and action. Previous iterations of EDI-WG surveys have resulted in guidelines for diversity-inclusive scientific meetings \cite{QuantumFlagship_EDI_2024}, which have since become the official policy of Quantum Flagship. 

In 2025 DiviQ and the EDI-WG conducted the largest international surveys to date on diversity, workplace climate, and opportunity in quantum science and technology. Over $1000$ participants—from students to senior specialists—contributed their perspectives. Their insights form the backbone of this paper and of our forthcoming full report. This work was supported by our partner organizations around the world, including We in Quantum Development (WiQD): a professional network for quantum industry, academia and policy supporting minoritized quantum scientists. 

Lastly, the two lead authors are associated with SPICE Science: an Oregon-based science camp that teaches middle school girls science, gives grad students first-hand experience with near-peer mentorship and science outreach, and conducts research into science identity formation \cite{Todd2017}.

\begin{figure}[t]
    \centering
    \includegraphics[width=\linewidth]{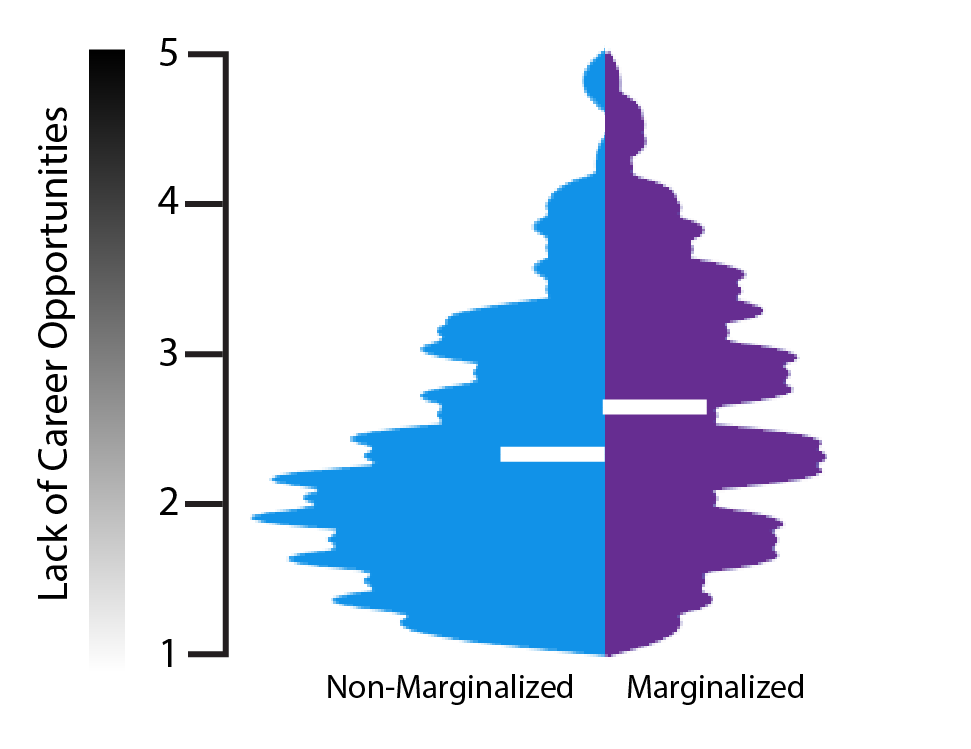}
                \caption{Mean differences in perception of career opportunities between non-marginalized ($\textit{M}=3.85, \,\textit{SD}=.85$), and marginalized respondents on a scale of 1-5 with 5 being the highest perception of opportunity ($\textit{\rm M}=3.41,\, \textit{SD}=.91 \,,\textit{t}[230]=3.82, \,\textit{p}<0.001$).}
    \label{fig:careeropps}
\end{figure}

\section{Research Principles}


Ethical examinations in quantum science often focus on technological, social, economic, and geopolitical impacts of quantum technology. Much of this discourse looks outward, asking how quantum innovation will transform the world. In this work, we propose and embrace an equally essential inward-facing view grounded in the foundational ethical frameworks of modern science: the Declaration of Helsinki and the Belmont Report \cite{DeclarationOfHelsinki1964,BelmontReport1979}. The principles at the heart of these documents offer a powerful guide for shaping the quantum workforce and the policies that govern it. They are part of the motivation for the style of survey we conducted (a community needs assessment) as well as the foundation for the framework through which we interpret our findings.

Greater attention to the work climate of quantum science and to the composition of its workforce is not only a good unto itself, but also a strategic advantage. Improving the retention of the quantum science workforce increases the size of that workforce. Furthermore, a more diverse community of quantum scientists improves the field’s capacity to anticipate, identify, and mitigate problems at the point of invention rather than after deployment. When researchers with a wide range of lived experiences—across gender, race, disability, culture, language, and identity—shape quantum technologies from the beginning, issues such as privacy, accessibility, bias, and social impact are surfaced earlier and more accurately. This leads to safer systems, stronger innovation pipelines, and more mature technologies. A diverse workforce will align with ethical, social, and economic priorities; a quantum workforce where people feel comfortable staying long-term strengthens both scientific integrity and global competitiveness.

\begin{figure}[b]
            \centering
            \includegraphics[width=\linewidth]{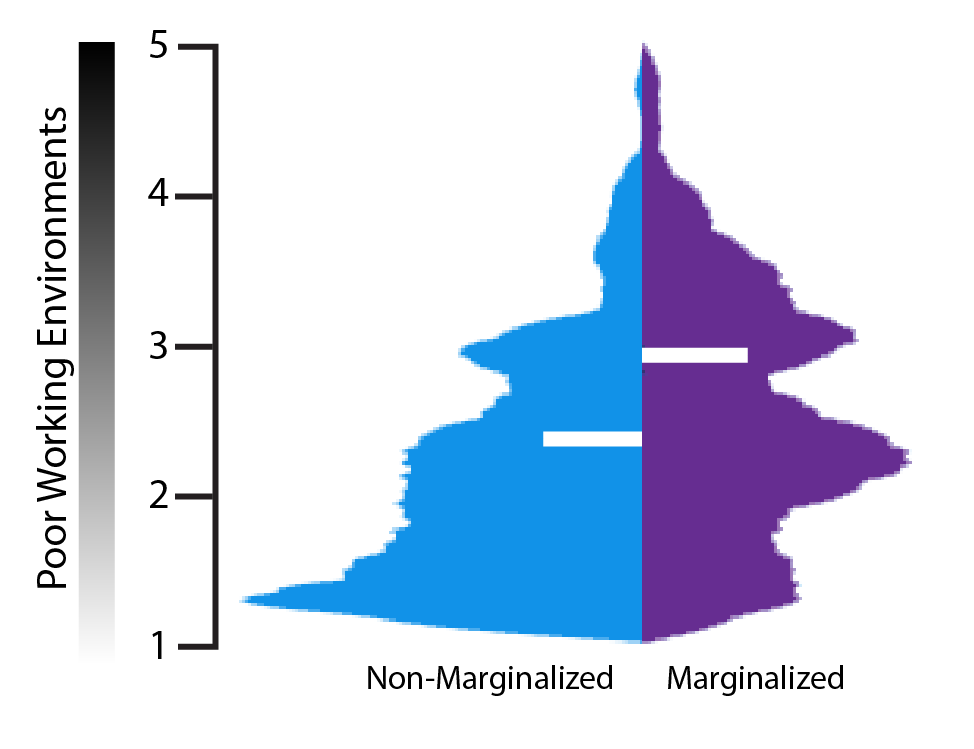}
            \caption{Mean differences in perception of working environment between non-marginalized ($\textit{M}=3.99, \,\textit{SD}=.85$), and marginalized respondents on a scale of 1-5 with 5 being the worst working environments ($\textit{ M}=3.54,\, \textit{SD}=.92 \,,\textit{t}[586]=6.06, \,\textit{p}<0.001$).}
            \label{fig:enviornment}
        \end{figure}
\lettersection{Respect for persons}In research with human-subjects, this principle protects the autonomy and dignity of research subjects. Applied to the quantum workforce, it demands that the field recognize and address the barriers that constrain full participation—whether they stem from discrimination, gatekeeping, hostile climates, or structural inequities. Respect means creating environments where researchers can contribute expertise without navigating disproportionate risk or exclusion; in short, valuing each person's unique way. This is especially important for researchers who are marginalized within their scientific community e.g. because of their race, gender, disability status, sexual orientation, ethnicity, or nationality. 


\lettersection{Beneficence} Beneficence requires minimizing harm and maximizing the benefits of research, literally ``good doing.'' Quantum science \emph{cannot} meet this standard if marginalized researchers—--particularly women, LGBTQ+ scientists, people with disabilities, and members of racialized groups—--face higher rates of harassment, bullying, and career obstacles. A field that generates cutting-edge technology must not generate preventable human cost. Investing in inclusive structures is therefore not ancillary; it is an ethical obligation.

\lettersection{Justice} Justice, in ethics, demands fair distribution of burdens and opportunities. In quantum, this translates to ensuring that training pathways, fellowships, internships, and research roles do not systematically privilege those with the most resources, the strongest networks, or the most conventional backgrounds. Justice requires multiple entry points—computer science, engineering, mathematics, chemistry, materials science, and beyond—as well as flexibility for those balancing caregiving, disability, migration, or nontraditional life trajectories.

Using these principles as a framing shifts diversity from a moral accessory to a scientific responsibility. Here, we endeavor to use the results of the $2025$ global quantum needs assessment survey to guide the construction of pathways towards alignment with these three principles. We believe the global quantum science community has an opportunity to embed these ethical commitments directly into workforce development, funding structures, and institutional expectations. 

\begin{figure}[t]
    \centering
    \includegraphics[width=\linewidth]{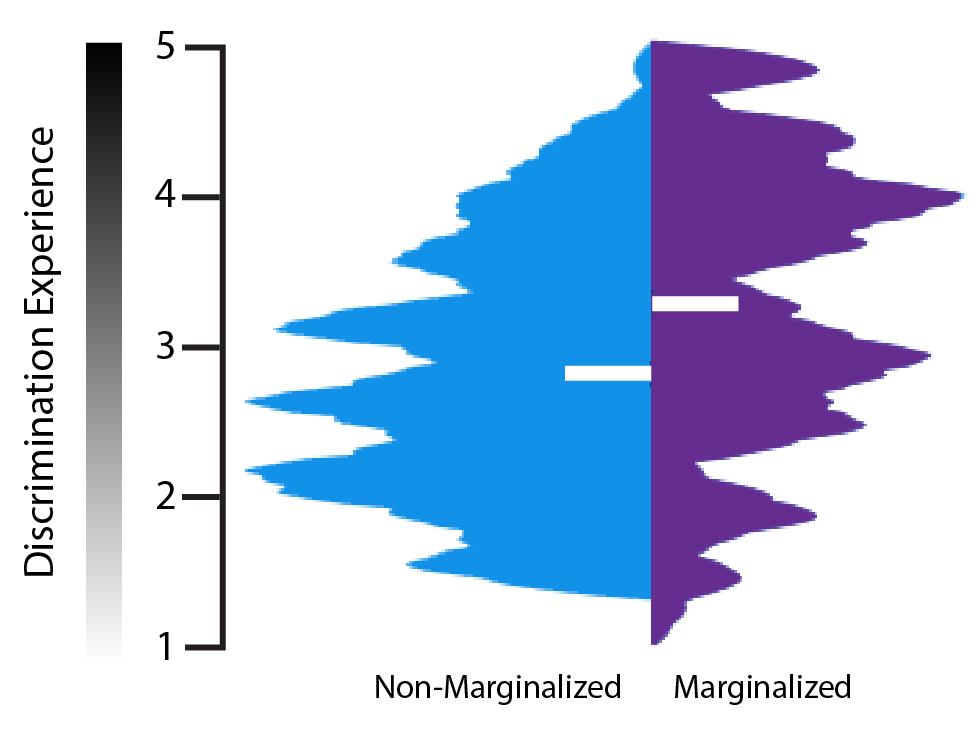}
    \caption{Mean differences in perception of prevalence of discrimination in quantum between non-marginalized ($\textit{M}=2.47, \,\textit{SD}=.91$), and marginalized respondents on a scale of 1-5 with 5 being high levels of discrimination ($\textit{ M}=3.35,\, \textit{SD}=1.02 \,,\textit{t}[358]=-8.60, \,\textit{p}<0.001$).}
    \label{fig:discrimination}
\end{figure}

\section{Early Findings From the Global Quantum Needs Assessment}

To support the development of an evidence-based understanding of equity and inclusion within the quantum science workforce, we conducted a preliminary climate assessment focused on experiences and perceptions of discrimination, bias, bullying, harassment, and workplace culture. Although quantum science is a rapidly expanding field with growing international investment and strategic importance, no comprehensive EDI-focused workforce assessment currently exists. This project represents an initial step toward filling that gap.

Across $1000+$ respondents—from undergraduate students to principle investigators (PIs), from academia to industry—a consistent message emerges: diversity, climate, and opportunity in quantum are recognized challenges across the field. The degree of concern, however, varies significantly depending on identity and career stage.
Four factors informing the state of the quantum workforce experience were examined: career viability, working environments, climate at meetings and events, prevalence of bias/discrimination as well as bullying/harassment. Respondents came from 36 different countries. $32\%$ percent of respondents identify as belonging to a marginalized group; $10\%$ percent identify as possessing a disability. Respondents spanned the career spectrum with $34\%$ identifying as interested, but not engaged in quantum science careers, $6.5\%$ working in a non-research career in quantum, $37.2\%$ were junior career scientists, $14.6\%$ were senior career scientists, and $2\%$ left a career in quantum science, and additional $6.2\%$ of respondents were uninterested in a career in quantum science.

\begin{figure}[t]
        \centering
        \includegraphics[width=\linewidth]{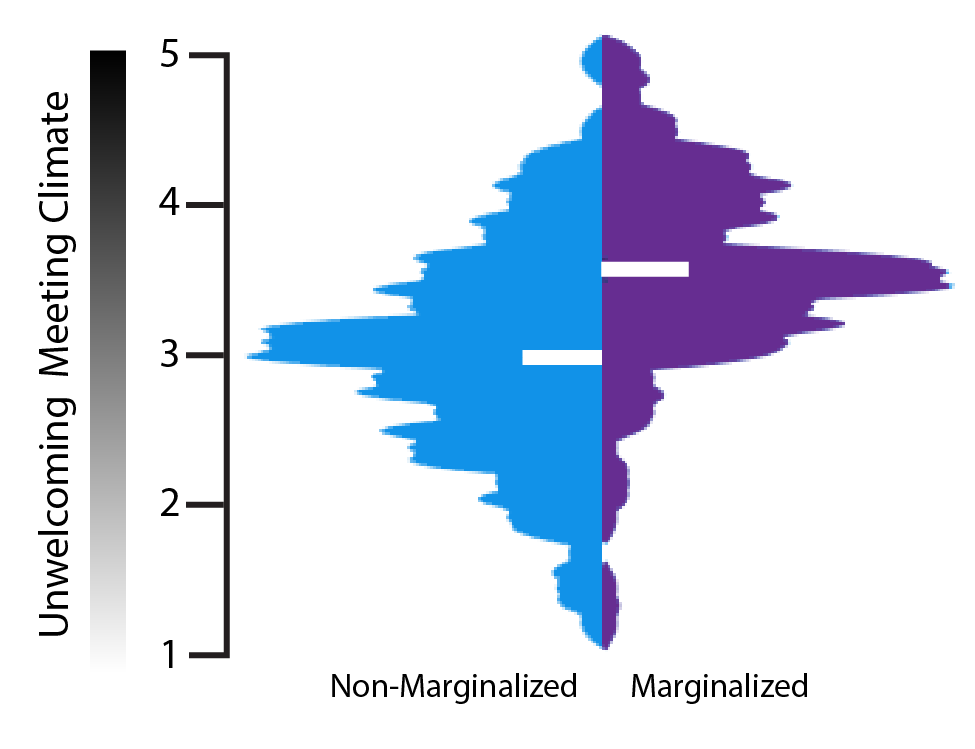}
        \caption{Mean differences in perception of climate at scientific meetings between non-marginalized ($\textit{M}=3.00, \,\textit{SD}=.77$), and marginalized respondents on a scale of 1-5 with 5 being the least welcoming climate ($\textit{ M}=3.51,\, \textit{SD}=.74 \,,\textit{t}[357]=-6.81, \,\textit{p}<0.001$).}
    \label{fig:career}
\end{figure}

\subsection{Marginalized Respondents Experience Quantum Science Differently}
Marginalized respondents consistently rated issues in all four factors more negatively to a statistically significant degree. These differences persist across career levels.  

 Figs. \ref{fig:leaving}-\ref{fig:discrimination} summarize responses to questions related to respondent experiences in and perceptions of quantum science as a discipline. Across the board, marginalized respondents reported poorer treatment, less recognition, and a lower likelihood of recommending their career or workplace to others. Notably, the majority of non-marginalized respondents also agreed that there are problems in all areas, though to a lesser degree among marginalized respondents.

Overall, perceptions of career opportunities in quantum and working environments are positive, but again, members of marginalized groups report a more negative view. Views of climate e.g. in workplaces and at scientific meetings, on the other hand, are consistently viewed as neutral or negative, with marginalized respondents perceiving issues of climate as poor and of more pressing concern.
Most tellingly, half of respondents from marginalized backgrounds report having considered leaving quantum science while only a small proportion of non-marginalized respondents report the same (Fig. \ref{fig:leaving}). Not only are marginalized quantum scientists experiencing greater hardships; those hardships are influencing their career decisions.
The picture painted by these results is grim from a workforce perspective. The general view of quantum careers is one of opportunity overbalanced by serious obstacles; obstacles that are perceived as greater by respondents from marginalized backgrounds (Figs. \ref{fig:careeropps}--\ref{fig:career}).

\begin{figure}
    \centering
    \includegraphics[width=\linewidth]{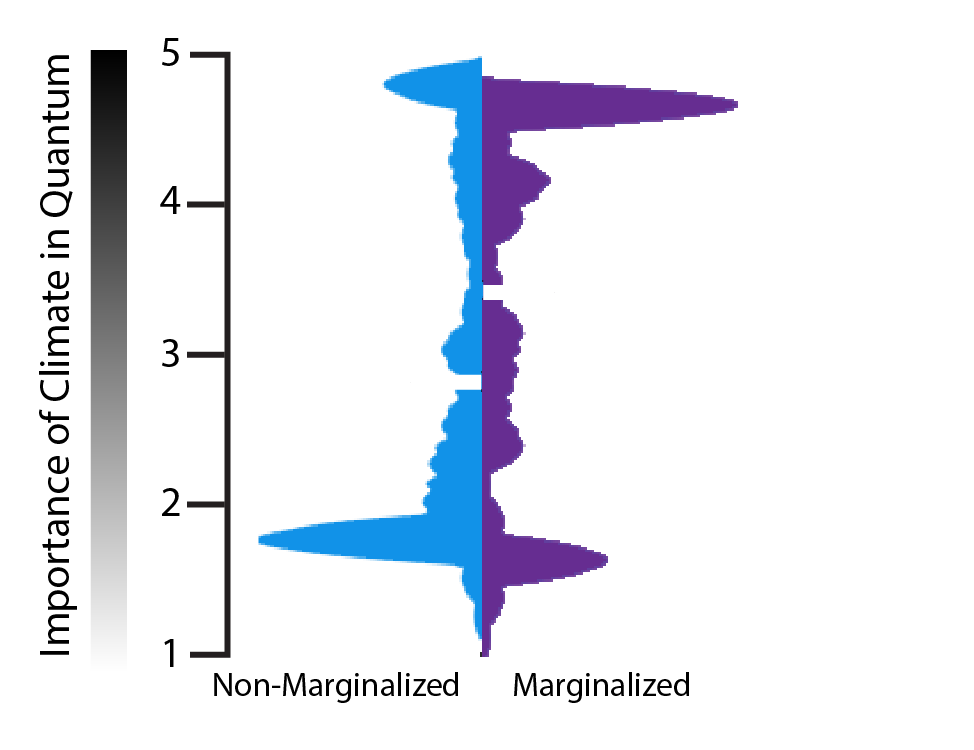}
    \caption{The mean differences between non-marginalized ($M=3.69,\, SD=.77$) and marginalized respondents in response to the importance of workplace, department, and ecosystem climate in quantum science on a scale of 1-5 with 5 being the most important. (${M}=4.19,\, \textit{SD}=.81,\, \textit{t}[218]=-4.60,\, \textit{p}<.001$).}
    \label{fig:climateissues}
\end{figure}

\subsection{Early Career Perceptions}
Respondents who are not part of the quantum workforce, but expressed an  \textit{interest} in quantum perceive opportunities for careers in quantum science but also challenges.  Respondents identified issues of inclusion, climate, protections from harassment, and preventing bias and discrimination as more pressing in quantum sciences and technology than in other disciplines. Fig. \ref{fig:climateissues} shows that both non-marginalized  and marginalized respondents  perceived the importance of these issues as high, though marginalized respondents perceive these issues as  more pressing than non-marginalized. Overall, respondents perceive quantum science as less diverse and less welcoming, especially for women and gender minorities; this perception itself is a barrier to entering the field.

\subsection{View from the Top}
Respondents at senior levels report that marginalized colleagues encounter more obstacles related to advancement, mentorship access, and workplace culture. Respondents overwhelmingly endorse greater accommodations and commitments to diversity with protections from bias and harassment ranking highest. Other practices viewed positively by respondents include recruitment policies that include diversity considerations, policies that support family commitments, equity and inclusion policies and the provision of facilities that accommodate all employees.  
Senior researchers remain hopeful about the future of diversity in quantum. 
\begin{itemize}
    \item  \textit{``I think the changes within the community have been positive, but still have a ways to go. When I began the community was nearly entirely straight, male, and white or asian, and there was little or no discussion of [diversity, equity, and inclusion] being an issue. The demographics of each new cohort seem to be more diverse along nearly all axes, and most (though not all!) of my senior colleagues recognize to some degree that (a) people from underrepresented groups face extra barriers and (b) the field is stronger when more people can contribute to it."}
\end{itemize}

While many senior researchers observed that the discipline has changed and diversified since they started out, they also acknowledged that, by and large, the people at the top are much the same in terms of demographics and the pathways they took to achieve their current status as they were decades ago. 
\begin{itemize}
\item  \textit{``The field has really grown in the last few years. With that there have been several initiatives to make the field more diverse and inclusive. While I think it's very nice overall that there is more awareness I don't think I've seen much change actually happening at higher levels. I hope this is a matter of time."}
\item \textit{``Things get names, that [sound] good, but people  and structures haven't changed."}
\end{itemize}

Concrete accommodations were praised as aids to advancement for marginalized quantum scientists.
\begin{itemize}
\item\textit{ ``I've seen, in a very positive light, how major conferences organized by APS or Optica offer childcare grants. As a mother of three children under the age of five, that has truly been a game changer for me. This kind of support should be promoted much more widely‚ almost to the point of being mandatory."}
\end{itemize}

Many senior respondents also cite the punishing nature of careers that are often viewed as intrinsically competitive, requiring 24/7 attention, and not friendly to family life. Several researchers called out work expectations they view as unsustainable and abusive.

\begin{itemize}
    \item \textit{``On the other hand, from a negative perspective, I believe the pressure for scientific productivity has skyrocketed, pushing a publish like there's no tomorrow, mentality‚ even at the cost of emotional or psychological abuse toward students and research teams."}
    \item \textit{``This CANNOT continue. Excellence in science cannot come at the expense of abuse. A truly excellent scientific environment must be built on respect, empathy, and care for the people who make science possible."}
    \item \textit{``I feel that there is more support from ground-up initiatives on this and voices of young scientists are better heard. However, I do still feel quite a lot of resistance coming from some parts of the community, mostly still having the fallacy that diversity (especially when it comes to hiring) and excellence are not attainable together."}
\end{itemize}
One senior researcher talked about their journey of understanding. 
\begin{itemize}
    \item \textit{``I had a very positive [experience] personally . . . I wasn't aware of how challenging it had been for some of my peers until I started discussing [equity and inclusion] more intentionally. It made me understand that these struggles and barriers can be easily missed or overlooked if we don't make an intentional effort to engage with others and discuss about them."}
\end{itemize}

\subsection{Mentorship}

Respondents from all levels express a desire for more structured mentorship, better institutional support for inclusive and nurturing supervision, and improved support networks e.g. formal and informal near-peer mentorship \cite{Akinla2018}. Below are selected quotes from the survey:
    \begin{itemize}
        \item \textit{``More built-up mentoring networks so that trainees can find advice, guidance, and advocacy from senior professionals beyond their supervisor and local lab/workplace"}
        \item \textit{``Near peer mentoring"}
    \end{itemize} 
    
    Respondents were not merely suggesting more mentoring for junior quantum scientists, but also resources and training for the senior scientists responsible for guiding them. Awareness is the first step, once aware however, many advisors feel challenged to support students from marginalized backgrounds.
    \begin{itemize}
        \item \textit{``I started my career essentially unaware of the extra barriers faced by people from underrepresented groups. I came to learn about it gradually and through a mix of influences: having a spouse (who works in a different field) describe the challenges she was facing; mentoring students from a wider range of groups and seeing their experiences; and more formal diversity training from the university."}
        \item \textit{``Physics always had a severe gender imbalance, finding causes and remedies was and is important. I have become more aware of the nuances and complexity of attracting and mentoring diverse students, as well as the barriers they face."}
        \item \textit{``More built-up mentoring networks so that trainees can find advisors beyond their supervisor and local lab/workplace.”}
    \end{itemize}

\subsection{The Ones Who Leave}
The number of respondents who left careers in quantum science were small (\textit{n=16}); however, their experiences are telling. They reported higher levels of bias and harassment. Nearly all cited issues of diversity, equity, and inclusion as parts of their decision to leave. One respondents summed their reasons for leaving succintly:
\begin{itemize}
\item \textit{``Bro-y and competitive culture."}
\end{itemize}
Another respondent called for institutions to prioritize collegiality and personal support as a means to improving success in quantum sciences.
\begin{itemize}

\item \textit{``[A] Broader desire by all professional organizations (academic, industrial, government) to value the contributions of individuals as human beings above all else. Making an impact on colleagues through collegiality for the betterment of team advancement and teaching/training/developing junior scientists/colleagues to be successful in the workplace are just as impactful work for all organizations as meeting profit goals or unfortunately far too often unrealistic/'stretch' targets promised to external parties"}
\end{itemize}

\begin{table*}[t!]
\caption{Comparison of our findings to negative interpersonal experiences reported astronomy, solar science, and geophysics reported in the Royal Astronomy Society (RAS) Report \cite{RAS2024bullying}, both for total sample populations and for marginalized populations.}\label{table:tbl}
\begin{ruledtabular}
\begin{tabular}{lcccc}
\textbf{Measure} 
& \textbf{Quantum Overall}& \textbf{Quantum: Marginalized } 
& \textbf{RAS } 
& \textbf{RAS: Marginalized} \\
\hline
Witnessed bias or discrimination   & 54\% & 73\%& --& -- \\
Experienced bias or discrimination & 39\% & 56\%& --   & -- \\
Witnessed bullying or harassment   & 38\% & 47\%& 57\% & ~65-75\%\\
Experienced bullying or harassment & 23\% & 28\%& 44\% & ~47-62\%\\
\end{tabular}
\end{ruledtabular}
\footnotetext{While our data required users to self-identify as marginalized/non-marginalized, in interpreting the RAS report we use the term marginalized to include racial/ethnic minorities, LGBTQ+ researchers, disabled researchers, and gender-diverse respondents.}
\end{table*}
\subsection{Pathways}
A common call at multiple levels, but particularly among junior scientists, is the need for more "on-ramps" into quantum science beyond the typical physics graduate student to professor model. Respondents want   diverse entry pathways into quantum careers, including smoother transitions from computer science, engineering, and mathematics. They want to see more support for interdisciplinary mobility.
\begin{itemize}

\item \textit{``There needs to be pathways back to quantum for people who have graduated in other areas." }

\item \textit{``Many talented researchers leave the field due to hostile environments or lack of mentorship, which represents a significant loss for the community. Creating structures that value diversity, ensure accountability, and provide clear career pathways would help retain talent and foster innovation in this rapidly evolving field.''}

\end{itemize}

Taken together, these findings suggest that the quantum ecosystem is losing talent early and creating avoidable friction late. Improving both the \textbf{realities} and \textbf{perceptions} of climate and opportunity will be essential if we aim to cultivate a robust, competitive, and ethically grounded quantum workforce.

\subsection{A Cross Discipline Comparison}

In $2023$, the UK Royal Astronomy Society released the results of their survey on bullying and harassment in the disciplines of astronomy, solar science, and geophysics. A comparison of the two studies show negative interpersonal experiences are prevalent and  disproportionately affect researchers with marginalized identities, with the findings summarized in Table 1. 


Although the absolute prevalence of direct victimization is higher in the RAS sample than in the present data ($44\%$ vs. $23\%$), both studies show that harmful behavior is widespread and highly visible within research environments. Furthermore, the two studies converge in their findings regarding inequitable impact; the RAS report documented that marginalized groups—including racial and ethnic minorities, LGBTQ+ researchers, disabled individuals, and gender-diverse respondents—experienced bullying and harassment at sharply elevated rates. Our present study reproduced this pattern; both junior and senior researchers who identified as marginalized reported significantly higher levels of witnessing and experiencing bias and bullying than their non-marginalized peers. 

Despite differences in discipline, geographic context, sampling, and measurement, the consistency of these patterns suggests that inequitable treatment of marginalized researchers reflects a broader structural problem across scientific cultures rather than isolated institutional issues. Both datasets underscore the need for organizational reforms that address not only individual incidents but the systems that enable biased or harmful behavior to persist.

\section{Conclusion: A Call to Action}

The quantum workforce of the future must be built on and embody the principles of responsible science: respect for persons, beneficence, and justice. Not only will these values ensure a responsible path forward, but in turn, they will help propel the quantum ecosystem through the direct benefits afforded by a diversified quantum community.

These values are not abstract. They have direct implications for hiring, training, funding, policy, and evaluation. Our survey of over $1000$ members of the quantum science community has identified key areas where progress can be made to improve alignment with these principles. Specifically, we call on policymakers, universities, and institutes to prioritize and allocate funding and resources to the following key areas:

\begin{itemize}    
    \item \textbf{Investment in and re-framing of diversity-hires explicitly as talent retention.} Improving senior staff representation from underrepresented communities, especially at the tenure track faculty level, will improve resiliency \cite{Barak2002}, combat systemic scholarly devaluation \cite{Settles2021}, and support new pathways for talent retention in the quantum workforce \cite{EC2005TurningDiversity,EC2008SMEDiversity}, in turn stimulating quantum research \cite{EC2008DiversityInnovation}. 
    \item \textbf{Mentorship, especially for junior researchers.} Both formal and informal mentorship will improve the quantum science community, especially for marginalized researchers \cite{colwell2020}. 
    \item \textbf{Grants to help cover additional costs} associated with caregiving responsibilities (e.g. traveling with children) and disability-related accommodations (e.g. accessible transportation, traveling with a personal aide), expanding on what is already offered \cite{IYQGlobalFund2025}, as well as childcare at conferences. 
    \item \textbf{Workshops, scholarships, and stable financial support to fund junior researchers}, especially for those from marginalized communities.
    \item \textbf{Network-forming and community support} so that students, especially those from marginalized communities, are able to access opportunities for career advancement across the globe. 
    \item \textbf{Requiring the development and enforcement of anti-harassment policies.} Improved training for senior staff and managers to both improve systems for monitoring progress and track accountability in implementing EDI measures. 
    \item \textbf{Investment in senior staff and managers to improve their EDI-competent mentorship capabilities.} This will amplify community-level and grassroots support for quantum diversity, especially by improving support for junior researchers as they overcome obstacles they encounter in their career.  
    \item \textbf{Continuing research into the needs of the global quantum science community.} This must continue in collaboration with social scientists to continue investigating how we can shift the culture of our field specifically to be actively welcoming to a diverse workforce. There is also a need for research into the economic impacts of diversity in the quantum workforce; a diverse quantum workforce will be more humane, more resilient, and more reflective of the societies it intends to serve, and also more productive. 
\end{itemize}

Acting on this call to action will support building a diverse, quantum-ready workforce. A quantum future that embodies the principles we've proposed will not only be stronger and more innovative--—it will be more humane, more resilient, and more reflective of the societies it intends to serve. 

Many of our authors are based in Europe, where the quantum ecosystem is at a crossroads. As the EU Commission shapes the Quantum Act, it has a rare chance to articulate not only a technical vision for the field, but an ethical and humane one that will also align with Europe's economic priorities. Notably, the European Union still lags behind the United States and China in terms of quantum investment \cite{QuIC2025}, and much of quantum science's culture is still dictated by other nations, especially the United States. However, the European Union still has the opportunity to be a frontrunner in quantum science by setting the right example in values. Mandates for quantum EDI directly from the European Commission, complimented by B-Corps \cite{BLab} style certification and diversity equality plans (similar to gender equality plans \cite{GEP}) along Horizon grant calls for diversity hires, would directly support the more local, community-specific efforts to improve EDI in quantum. This provides viable paths toward implementation of the key areas highlighted above. 

DiviQ and the Quantum Flagship EDI-WG stand ready to support this---along with similar efforts throughout the global quantum ecosystem---through evidence-based recommendations, community engagement, collaboration with our partner organizations around the world, and the forthcoming full report from our global survey. We also recognize that we, the authors of this report, have the privilege of being able to openly support quantum diversity efforts---not all quantum science researchers are so lucky; DiviQ stands ready to support individuals and grassroots organizations around the world in building quantum resiliency in the face of anti-EDI repression. 

As we finalize our initial data collection, which is available along with printable promotional material online on the DiviQ and Quantum Flagship websites \cite{DiviQSurvey,QuantumFlagshipSurvey}, we also emphasize that we are still far from our goal of $2000$ responses to the survey by the end of $2026$. We hope the open source data we have gathered, available online as well \cite{kaggle2025}, will continue to grow and be used to support the global quantum science ecosystem.

\section{Acknowledgments}

This work was partially funded by CEX2024-001490-S [MICIU/AEI/10.13039/501100011033], Fundació Cellex, Fundació Mir-Puig, and Generalitat de Catalunya through CERCA.
TzBP gratefully acknowledges support from the Quantum Software Consortium Ada Lovelace Fellowship. 

The survey itself was funded by Quantum Flagship and donations from SandboxAQ. 

We also gratefully acknowledge logistical support in survey distribution from the American Physical Society, We in Quantum Development (WiQD), Women for Quantum (W$4$Q, Attom*innen, the Quantum Internet Alliance, Quantum Software Consortium, QuCats, International Women$+$ in Theoretical Physics, the other members of the Quantum Flagship EDI Working Group (EDI-WG), and the individuals who brought our survey posters to their home institutions around the world. We are especially grateful to W$4$Q member Christiane Koch for feedback on an early versions of the survey, and EDI-WG member Irene D'Amico for feedback on tnis report. We are also grateful to the other members of the Quantum Flagship EDI Working Group, Diversity in Quantum, and our quantum diversity partner organizations for their ongoing support with this project.  Lastly, thank you to the over $1000$ survey respondents (so far) who have been put their time into helping us build towards a better quantum future.

\bibliography{globalneeds}

\appendix

\section{Methodology}

\subsection{\textbf{Participants and Recruitment}}
Respondents were recruited through a combination of professional networks, snowball sampling, and—most substantively—direct outreach at major scientific conferences, including the APS March Meeting. Participation was voluntary and open to individuals working or training in quantum-adjacent fields (e.g., physics, engineering, computer science, and industry roles connected to quantum technologies). Because the assessment was exploratory and field-wide sampling frames do not yet exist, the resulting sample should be interpreted as a convenience sample rather than a representative cross-section of the workforce.
\subsection{\textbf{Survey Instrument}}
The survey included a set of Likert-type items (1 = strongly disagree to 5 = strongly agree) designed to measure perceptions of climate, equity, diversity, inclusion, bias/discrimination, and bullying/harassment. Items were organized into thematic clusters reflecting conceptual constructs identified in the EDI literature and through consultation with the research team. Respondents were also invited to provide open-ended comments describing their experiences and recommendations for improving the quantum workforce climate.
\textbf{
\subsection{Analytic Approach}}
 Quantitative analyses included confirmatory factor analysis (CFA) to assess whether items grouped within the same construct demonstrated adequate communality and coherence. The CFA supported the proposed factor structure, indicating that the thematic item sets measured underlying shared concepts as intended. Group differences were examined using independent-samples \textit{t}-tests comparing marginalized and non-marginalized respondents across discrimination, bias, and bullying/harassment measures—an equity-centered framing aligned with the aims of the project. Qualitative responses were reviewed and coded thematically to contextualize quantitative findings and identify recurring patterns not captured by closed-ended items.
\subsection{ \textbf{Status of Findings}}
Because this is a position paper based on an initial round of data collection, results should be considered preliminary. Additional rounds of recruitment, instrument refinement, and deeper statistical modeling are planned to develop a comprehensive, field-wide assessment of equity and inclusion in the quantum workforce.

\subsection{Limitations}

These findings should be interpreted within the context of several limitations inherent to this preliminary assessment. First, the sample was not randomly drawn from the quantum science workforce; recruitment relied on convenience sampling, professional networks, and in-person outreach at major conferences, all of which conducted in English. As a result, the dataset may over-represent individuals who are more professionally engaged, more connected to EDI efforts, more motivated to share their experiences, and more comfortable answering a survey in English. Although the statistical tests conducted yielded large \textit{t}-values (ranging from approximately 4 to 9), these values reflect differences within this sample; concrete generalization to the full quantum workforce requires replication with a more systematically sampled population.

Second, because standardised workforce demographic baselines do not yet exist for quantum science, it is not possible to evaluate representativeness of this survey across marginalized identities, career stages, or institutional sectors. Relatedly, the use of broad categorical groupings (e.g., “marginalized” and “non-marginalized”) limits the ability to examine intersectional variation across specific identities.

Finally, the survey instrument is still under development. Although confirmatory factor analysis supported the coherence of the thematic constructs, additional validation work, longitudinal data collection, and iterative refinement will be necessary to fully establish the reliability and generalizability of these measures.

Despite these limitations, the consistency and magnitude of observed group differences, combined with qualitative accounts, point to meaningful and actionable concerns warranting both action and further investigation, as emphasised in the maintext.

\end{document}